\documentclass[aps, pra, twocolumn,showpacs,preprintnumbers,amsmath,amssymb]{revtex4}

\newcommand{\ket}[1]{\mbox{$ | #1 \rangle $}}
\newcommand{\bra}[1]{\mbox{$ \langle #1 | $}}
\usepackage[dvips]{graphicx}

\begin{document}

\preprint{}

\title{Multiparty controlled quantum secure direct communication using
Greenberger-Horne-Zeilinger state}

\author{Jian Wang}

 \email{jwang@nudt.edu.cn}

\affiliation{School of Electronic Science and Engineering,
\\National University of Defense Technology, Changsha, 410073, China }
\author{Quan Zhang}
\affiliation{School of Electronic Science and Engineering,
\\National University of Defense Technology, Changsha, 410073, China }
\author{Chao-jing Tang}
\affiliation{School of Electronic Science and Engineering,
\\National University of Defense Technology, Changsha, 410073, China }


\begin{abstract}
Base on the idea of dense coding of three-photon entangled state and
qubit transmission in blocks, we present a multiparty controlled
quantum secret direct communication scheme using
Greenberger-Horne-Zeilinger state. In the present scheme, the sender
transmits her three bits of secret message to the receiver directly
and the secret message can only be recovered by the receiver under
the permission of all the controllers. All three-photon entangled
states are used to transmit the secret messages except those chosen
for eavesdropping check and the present scheme has a high source
capacity because Greenberger-Horne-Zeilinger state forms a large
Hilbert space.
\end{abstract}

\pacs{03.67.Dd, 03.67.Hk}
\keywords{Quantum key distribution; Quantum teleportation}
\maketitle


%
%
Quantum communication has been one of the most promising
applications of quantum information science. Quantum key
distribution (QKD) which provides unconditionally secure key
exchange has progressed quickly since the first QKD protocol was
proposed by Benneett and Brassard in 1984 \cite{bb84}. A good many
of other quantum communication schemes have also been proposed and
pursued, such as Quantum secret sharing
(QSS)\cite{hbb99,kki99,zhang,gg03,zlm05,xldp04}, quantum secure
direct communication (QSDC)
\cite{beige,Bostrom,Deng,denglong,cai1,cai4,jwang,cw1,cw2,tg,zjz}.
QSS is the generalization of classical secret sharing to quantum
scenario and can share both classical and quantum messages among
sharers. Many researches have been carried out in both theoretical
and experimental aspects after the pioneering QSS scheme proposed
by Hillery, Buz\v{e}k and Berthiaume in 1999 \cite{hbb99}.
Different from QKD, QSDC's object is to transmit the secret
messages directly without first establishing a key to encrypt
them. QSDC can be used in some special environments which has been
shown by Bostr\"{o}em and Deng et al.\cite{Bostrom,Deng}. The
works on QSDC attracted a great deal of attentions and can be
divided into two kinds, one utilizes single photon
\cite{denglong,cai1}, the other utilizes entangled state
\cite{Bostrom,Deng,cai1,cai4,jwang,cw1,cw2,tg,zjz}. Deng et al.
proposed a QSDC scheme using batches of single photons which
serves as one-time pad \cite{denglong}. Cai et al. presented a
deterministic secure direct communication scheme using single
qubit in a mixed state \cite{cai1}. The QSDC scheme using
entanglement state is certainly the mainstream. Bostr\"{o}m and
Felbinger proposed a "Ping-Pong" QSDC protocol which is
quasi-secure for secure direct communication if perfect quantum
channel is used \cite{Bostrom}. Cai et al. pointed out that the
"Ping-Pong" Protocol is vulnerable to denial of service attack or
joint horse attack with invisible photon \cite{cai2,cai3}. They
also presented an improved protocol which doubled the capacity of
the "Ping-Pong" protocol \cite{cai4}. Deng et al. put forward a
two-step QSDC protocol using Einstein-Podolsky-Rosen (EPR) pairs
\cite{Deng}. We presented a QSDC scheme using EPR pairs and
teleportation \cite{jwang}. Chuan Wang et al. proposed a QSDC
scheme with quantum superdense coding \cite{cw1} and a multi-step
QSDC scheme using GHZ state \cite{cw2}. Ting Gao et al. and
Zhan-jun Zhang et al. each presented a QSDC scheme using
entanglement swapping \cite{tg,zjz}.

In this paper, we present a multiparty controlled QSDC (MCQSDC)
scheme using GHZ state and its transformation. In the present
scheme, the sender's secret message is transmitted directly to the
receiver and can only be reconstructed by the receiver with the help
of all the controllers. Different to QSS, the sender transmits his
or her secret message to the receiver directly and the information
of the receiver is asymmetric to those of the controllers. Our
scheme employs dense coding of three-photon entangled state proposed
by H. J. Lee et al.\cite{lee} and qubit transmission in
blocks\cite{long}. Eight possible states of GHZ state which form a
complete orthonormal basis, carry three bits of information and all
GHZ states are used to transmit the secret messages except those
chosen for eavesdropping check. We also discuss the security of the
scheme, which is unconditionally secure.


We first present a single party controlled QSDC scheme (CQSDC) and
then generalize it to a MCQSDC scheme. In the scheme, we utilize
dense coding of three-photon GHZ state. There are eight independent
three-photon GHZ states which form a complete orthonormal basis,
namely
\begin{eqnarray}
\label{1}
\ket{\Psi_1}=\frac{1}{\sqrt{2}}(\ket{000}+\ket{111})=\frac{1}{\sqrt{2}}(\ket{\phi^+}\ket{+}+\ket{\phi^-}\ket{-}),\nonumber\\
\ket{\Psi_2}=\frac{1}{\sqrt{2}}(\ket{000}-\ket{111})=\frac{1}{\sqrt{2}}(\ket{\phi^-}\ket{+}+\ket{\phi^+}\ket{-}),\nonumber\\
\ket{\Psi_3}=\frac{1}{\sqrt{2}}(\ket{100}+\ket{011})=\frac{1}{\sqrt{2}}(\ket{\psi^+}\ket{+}-\ket{\psi^-}\ket{-}),\nonumber\\
\ket{\Psi_4}=\frac{1}{\sqrt{2}}(\ket{100}-\ket{011})=\frac{1}{\sqrt{2}}(\ket{\psi^+}\ket{-}-\ket{\psi^-}\ket{+}),\nonumber
\end{eqnarray}
\begin{eqnarray}
\ket{\Psi_5}=\frac{1}{\sqrt{2}}(\ket{010}+\ket{101})=\frac{1}{\sqrt{2}}(\ket{\psi^+}\ket{+}+\ket{\psi^-}\ket{-}),\nonumber\\
\ket{\Psi_6}=\frac{1}{\sqrt{2}}(\ket{010}-\ket{101})=\frac{1}{\sqrt{2}}(\ket{\psi^+}\ket{-}+\ket{\psi^-}\ket{+}),\nonumber\\
\ket{\Psi_7}=\frac{1}{\sqrt{2}}(\ket{110}+\ket{001})=\frac{1}{\sqrt{2}}(\ket{\phi^+}\ket{+}-\ket{\phi^-}\ket{-}),\nonumber\\
\ket{\Psi_8}=\frac{1}{\sqrt{2}}(\ket{110}-\ket{001})=\frac{1}{\sqrt{2}}(\ket{\phi^+}\ket{-}-\ket{\phi^-}\ket{+}),\nonumber\\
\end{eqnarray}
where
\begin{eqnarray}
\ket{+}=\frac{1}{\sqrt{2}}(\ket{0}+\ket{1}),\ket{-}=\frac{1}{\sqrt{2}}(\ket{0}-\ket{1}),\nonumber\\
\ket{\phi^+}=\frac{1}{\sqrt{2}}(\ket{00}+\ket{11}),\ket{\phi^-}=\frac{1}{\sqrt{2}}(\ket{00}-\ket{11}),\nonumber\\
\ket{\psi^+}=\frac{1}{\sqrt{2}}(\ket{01}+\ket{10}),\ket{\psi^-}=\frac{1}{\sqrt{2}}(\ket{01}-\ket{10}).
\end{eqnarray}
$\ket{\Psi_k}, k=1,2,\cdots,8$ can be transformed into each other by
performing one of four unitary operations,
\begin{eqnarray}
I=\ket{0}\bra{0}+\ket{1}\bra{1},\nonumber\\
\sigma_z=\ket{0}\bra{0}-\ket{1}\bra{1},\nonumber\\
\sigma_x=\ket{0}\bra{1}+\ket{1}\bra{0},\nonumber\\
i\sigma_y=\ket{0}\bra{0}-\ket{1}\bra{1},
\end{eqnarray}
on any two of the three photons. Suppose the initial state is
$\ket{\Psi_1}$, we illustrate the transformation of GHZ states in
Table 1.
\begin{table}[h]
\caption{The transformation of GHZ states by performing operations
on two photons }\label{Tab:one}
  \centering
    \begin{tabular}[b]{|c|c|} \hline
       & unitary operations on the first and the second photon\\ \hline
      \ \ket{\Psi_1} & $\sigma_z\otimes\sigma_z$   or $I\otimes I$ \\ \hline
      \ \ket{\Psi_2} & $I\otimes\sigma_z$    or $\sigma_z\otimes I$ \\ \hline
      \ \ket{\Psi_3} & $i\sigma_y\otimes\sigma_z$   or $\sigma_x\otimes I$ \\ \hline
      \ \ket{\Psi_4} & $\sigma_x\otimes\sigma_z$    or $i\sigma_y\otimes I$ \\ \hline
      \ \ket{\Psi_5} & $I\otimes\sigma_x$      or $\sigma_z\otimes i\sigma_y$ \\ \hline
      \ \ket{\Psi_6} & $\sigma_z\otimes\sigma_x$  or $I\otimes i\sigma_y$ \\ \hline
      \ \ket{\Psi_7} & $\sigma_x\otimes\sigma_x$  or $i\sigma_y\otimes i\sigma_y$ \\ \hline
      \ \ket{\Psi_8} & $i\sigma_y\otimes\sigma_x$ or $\sigma_x\otimes i\sigma_y$ \\ \hline
    \end{tabular}
\end{table}

In order to distinguish the sender's operations correctly in our
scheme, we select eight two-photon operations, $U_1, U_2,\cdots,U_8$
from sixteen operations, where
\begin{eqnarray}
  &U_1=\sigma_z\otimes\sigma_z, &U_2=I\otimes\sigma_z, \nonumber\\
  &U_3=i\sigma_y\otimes\sigma_z,&U_4=\sigma_x\otimes\sigma_z, \nonumber\\
  &U_5=I\otimes\sigma_x, &U_6=\sigma_z\otimes\sigma_x,\nonumber\\
  &U_7=\sigma_x\otimes\sigma_x,&U_8=i\sigma_y\otimes\sigma_x.
\end{eqnarray}
Thus we obtain $\ket{\Psi_k}$ ($k=1,\cdots,8$) if $U_k$ is performed
on the $\ket{\Psi_1}$. We now describe the CQSDC scheme in detail.
Suppose the sender Alice want to transmit her secret message
directly to the receiver Charlie under the control of the controller
Bob.

(S1) Bob prepares an ordered $N$ three-photon states. Each of the
three-photon states is in the state
\begin{eqnarray}
\ket{\Psi}=\frac{1}{\sqrt{2}}(\ket{000}+\ket{111})_{ABC}.
\end{eqnarray}
We denotes the ordered $N$ three-photon qubits with
\{[P$_1(A)$,P$_1(B)$,P$_1(C)$], [P$_2(A)$,P$_2(B)$,P$_2(C)$],
$\cdots$, [P$_N(A)$,P$_N(B)$,P$_N(C)$]\}, where the subscript
indicates the order of each three-photon in the sequence, and $A$,
$B$, $C$ represents the three photons of each state, respectively.
Bob takes one particle from each state to form an ordered partner
photon sequence [P$_1(A)$, P$_2(A)$,$\cdots$, P$_N(A)$], called $A$
sequence. The remaining partner photons compose $B$ sequence,
[P$_1(B)$, P$_2(B)$,$\cdots$, P$_N(B)$] and $C$ sequence, [P$_1(C)$,
P$_2(C)$,$\cdots$, P$_N(C)$]. Bob selects randomly one of four
unitary operations, $I, \sigma_z, \sigma_x, i\sigma_y$ and performs
it on each of the photons in the $B$ sequence. He then sends the $A$
sequence and the $B$ sequence to Alice and keeps the $C$ sequence.

(S2) After receiving the $A$, $B$ sequence, Alice selects randomly a
sufficiently large subset from $A$, $B$ sequence for eavesdropping
check.

The procedure of the eavesdropping check is as follows: (a) Alice
announces publicly the positions of the selected photons. (b) Bob
publishes his operations which performed on the sampling photons in
the $B$ sequence. (c) Alice then chooses randomly a measuring basis
$Z$-basis(\ket{0},\ket{1}) or Bell-basis (\ket{\phi^\pm},
\ket{\psi^\pm}) to measure the selected photons in the $A$, $B$
sequence and announces publicly the measuring basis for each of the
sampling photons. (d) If Alice performs a $Z$-basis measurement, Bob
also performs $Z$-basis measurement on the corresponding photons in
the $C$ sequence; If Alice performs a Bell-basis measurement, Bob
performs a $X$-basis (\ket{+}, \ket{-}) measurement. After
measurements, Bob publishes his measurement results. Because of
Bob's unitary operations, $\ket{\Psi}$ is changed to one of four GHZ
states \ket{\Psi_1}, \ket{\Psi_2}, \ket{\Psi_5}, \ket{\Psi_6}.
According to Eq.\ref{1}, Alice and Bob can check the existence of
eavesdropper by comparing their measurement results. If the error
rate exceeds the threshold, they have to abort the communication.
Otherwise they continue to execute the next step.

(S3) Bob chooses randomly one of four unitary operations, $I,
\sigma_z, \sigma_x, i\sigma_y$ and performs it on each of the
photons in the $C$ sequence. He then sends the $C$ sequence photons
to Charlie.

(S4) Alice and Charlie analyze the error rate of the transmission of
$C$ sequence. The method of eavesdropping check is similar to that
of the step 2, but Bob should announce his operations on the
selected photons in the $C$ sequence after Charlie publish his
measurement results. If the error rate is below the threshold, they
proceed to execute the next step. Otherwise they abort the
communication.

(S5) Alice first selects randomly two sufficiently large subsets
from $A$, $B$ sequence for eavesdropping checks. She then performs
randomly one of eight operations, $U_1, U_2, \cdots,U_8$ on the
sample photons of $A$, $B$ sequence. The operation $U_k$
($k=1,\cdots\,8$) executed on the $A$, $B$ sequence makes the state
$\ket{\Psi}$ become one of the eight independent GHZ states,
$\ket{\Psi_1}$, \ket{\Psi_2}, $\cdots$, $\ket{\Psi_8}$. Alice, Bob
and Charlie agree that the eight GHZ states \ket{\Psi_1},
\ket{\Psi_2}, $\cdots$, \ket{\Psi_8}, represent a 3-bit of secret
message, that is \ket{\Psi_1}$\rightarrow$000,
\ket{\Psi_2}$\rightarrow$001, \ket{\Psi_3}$\rightarrow$010,
\ket{\Psi_4}$\rightarrow$011, \ket{\Psi_5}$\rightarrow$100,
\ket{\Psi_6}$\rightarrow$101, \ket{\Psi_7}$\rightarrow$110,
\ket{\Psi_8}$\rightarrow$111. Alice encodes her secret messages on
the remaining photons of $A$, $B$ sequence by performing the unitary
operations $U_k(k=1,\cdots,8)$. She then sends the $B$ sequence to
Charlie. After hearing from Charlie, Alice measures the sampling
photons of one subset of $A$ sequence in $Z$-basis or $X$-basis
randomly. She then publishes the positions of the sampling photons
and the measuring basis to each of the sampling photons. If Alice
performs a $Z$-basis ($X$-basis) measurement, Charlie performs a
$Z$-basis (Bell-basis) measurement on the corresponding sampling
photon in the $B$, $C$ sequence. After measurements, he publishes
his measurement results. Alice then lets Bob announce his operations
on the sampling photons of $B$, $C$ sequence. Thus Alice and Charlie
can check eavesdropping by comparing their measurement results. If
they make certain that there is no eavesdropper, Alice sends the $A$
sequence to Charlie. After hearing from Charlie, Alice publishes the
positions of the sampling photons of the other subset and lets
Charlie make GHZ basis measurements on the sampling photons of $A$,
$B$, $C$ sequence. After Charlie announces his measurements results,
Alice lets Bob publish his operations on the sampling photons in the
$B$, $C$ sequence. They then estimate the error rate of the
transmission of $A$ sequence. If there is no eavesdropping, they
continue to the next step, otherwise they abandon the communication.

(S6) Thus Charlie owns the $A$, $B$, $C$ sequence. Without Bob's
permission, Charlie can not obtain Alice's secret messages. Only
after Bob published his operations on the photons of $B$, $C$
sequence, could Charlie acquire Alice's secret messages by
performing a GHZ basis measurement on each of the GHZ states.

So far we have presented the CQSDC scheme. The security of the
present scheme is similar to those in Refs. \cite{Deng,long}. The
participators check the existence of eavesdropper during each
transmission of photon sequence. The security for the transmission
of photons can be reduced to the security of the BBM92 protocol
\cite{bbm92}. The process for the transmission of $A$, $B$ sequence
from Bob to Alice is similar to that in Ref. \cite{hbb99}. Alice and
Bob measure the sampling photons in a randomly selected measuring
basis, which ensures the security of the transmission of photon
sequence. Bob first performs a random unitary operation on the $B$
sequence and then sends it to Alice, which prevent Charlie from
acquiring partial secret message without the permission of Bob.
Without Bob's operations, Charlie performs GHZ basis measurement on
photons in the $A$, $B$, $C$ sequence and she can obtain partial
information of Alice at the step 6 of the scheme. The security for
the transmission of $C$ sequence in our scheme is similar to that of
the $C$ sequence in Deng et al.'s two-step QSDC protocol
\cite{Deng}. The transmission of $A$, $B$ sequence from Alice to
Charlie is the same as the transmission of $M$ sequence in the two
step protocol. Because of the qubit correlation of each GHZ state,
eavesdropper's eavesdropping will be detected during the
eavesdropping check. Charlie cannot obtain Alice's secret without
Bob's permission because Bob performs random unitary operations on
the photons in the $B$, $C$ sequence. Thus the present scheme is
unconditional security.

We then generalize the CQSDC scheme to a MCQSDC one. Suppose Alice
is the sender; Bob, Charlie, Dick,$\cdots$, York are the
controllers; Zach is the receiver. The first two step of the MCQSDC
scheme is the same as those of CQSDC scheme. We describe the MCQSDC
scheme from the step 3.

(S3$'$) Bob chooses randomly the Hadamard operation,
\begin{eqnarray}
H=\frac{1}{\sqrt{2}}(\ket{0}\bra{0}-\ket{1}\bra{1}+\ket{0}\bra{1}+\ket{1}\bra{0})
\end{eqnarray}
or the identity operation. He also chooses randomly one of four
unitary operations, $I, \sigma_z, \sigma_x, i\sigma_y$. Bob then
performs these two operations on each of the photons in the $C$
sequence. He then sends the $C$ sequence photons to Charlie.

(S4$'$) After receiving the $C$ sequence, Charlie performs the
similar operations as Bob and sends it to the next controller, Dick.
Dick and the remaining controllers repeat the similar operations as
Charlie until the receiver, Zach receives the $C$ sequence. The $H$
operation is very important for the security of the scheme. Suppose
the participators only performs $I$, $\sigma_z$, $\sigma_x$ or
$i\sigma_y$ operations on the $C$ sequence randomly and an
eavesdropper, Eve intercepts the $C$ sequence. She then prepares a
fake $C$ sequence which belongs to one part of EPR pairs and sends
it to the next participator, say Charlie. Eve can also intercepts
the sequence on which Charlie performed his operations. She then
performs Bell basis measurement on the intercepted sequence and her
other part of EPR pairs. Thus Eve can acquire the operation
information of the controller Charlie.

(S5$'$) After hearing from Zach, Alice selects randomly a
sufficiently large subset from $A$, $B$ sequence to check
eavesdropping. She then publishes the position of sampling photons.
She lets Zach measure the sampling photons in the $C$ sequence by
using either $Z$-basis or $X$-basis and publishes his measurement
results. For each of the sampling photon, Alice randomly selects a
controller to announce his operation information firstly and then
the others publish their operation information on the sampling
photons in turn. Suppose the number of $H$ operations performed on
each sampling photon in the $C$ sequence by the controllers is odd.
If Alice lets Zach measure the sampling photon in the $Z$-basis
($X$-basis), she performs Bell basis ($Z$-basis) measurement on her
corresponding two photons. Suppose the number of $H$ operations is
even. If Zach performs $Z$-basis ($X$-basis) measurement, Alice
measures the corresponding photons in $Z$-basis (Bell basis). After
doing these, Alice can determine the error rate of the transmission
of $C$ sequence. If she confirms there is no eavesdropping, the
process is continued. Otherwise, the process is stopped.

(S6$'$) Alice chooses the subsets of the sampling photons, encodes
her secret messages on the $AB$ sequence and transmits the $A$,
$B$ sequence to Zach step by step in the same way as the step 5 of
the CQSDC scheme. During the eavesdropping checks for the
transmission of $A$ sequence, Alice lets the controllers Bob,
Charlie, $\cdots$, York announce their operations on the sampling
photons. If the number of $H$ operations is odd, Zach first
performs $H$ operation on each of the selected photon in the $C$
sequence and then measures each of the corresponding three-photon
in GHZ basis. If the number of $H$ operations is even, Zach
performs GHZ basis measurement directly. Alice then lets Zach
publish his measurement results. Thus Alice and Charlie can
estimate the error rate of the transmission of $A$ sequence.
Actually, Eve can only interrupt the transmission of $A$, $B$
sequence and cannot steal any information even if she attacks the
photons.

(S7$'$) If the controllers permit Zach to reconstruct Alice's secret
messages, they tell Zach their operation information. If the number
of $H$ operations is odd, Zach first performs $H$ operation on the
corresponding photon and then measure the three-photon in GHZ basis.
If the number of $H$ operations is even, Zach performs GHZ basis
measurement directly. Thus Zach can obtain Alice's secret messages
under the permission of the controller Bob, Charlie, $\cdots$, York.

The $H$ operations performed by controllers can prevent Eve or a
dishonest controller from obtaining the control information, which
we have described it in the step 4$'$. During the eavesdropping
check for the transmission of $C$ sequence, Alice fist lets Zach
publish his measurement results and then the controllers announce
their operation information, which ensure each controller can really
act as a controller. If the controllers first publish their
operation information, Zach can obtain Alice's secret as long as he
acquires Bob's permission. In the present scheme, only with the
permissions of all the controllers could Zach acquire the secret
messages. The security of MCQSDC scheme is the same as that of
CQSDC.

So far we have presented a MCQSDC scheme based on dense coding of
three-photon entangled state and qubit transmission in blocks. We
first present a CQSDC scheme controlled by a single controller, and
then generalize it to a MCQSDC scheme. The sender encodes three bits
of secret message on two photons of GHZ state and the receiver can
reconstruct the sender's secret with the permissions of all the
controllers. It seems like a QSS scheme, but the sender's secret can
send to the receiver directly and the information of the receiver is
asymmetric to the controllers. We also analyzed that the security of
this scheme is the same as that of the BBM92 protocol, which is
unconditionally secure. In the present scheme, all of the GHZ states
are used to transmit the secret except those used to check
eavesdropping. The scheme has a high source capacity in that GHZ
state forms a large Hilbert space.



\begin{acknowledgments}
This work is supported by the National Natural Science Foundation of
China under Grant No. 60472032.
\end{acknowledgments}

%
%

%
%
\end{document}